%% file: main.tex
\documentclass[a4paper, oneside, twocolumn, notitlepage, 10pt]{extarticle_ecoc}

\usepackage{ecoc}
\usepackage[mode=tex]{standalone} 
\usepackage{subfigure}
\usepackage{dsfont}
\usepackage{float}
\usepackage{algorithm}
\usepackage[noend]{algpseudocode}
\usepackage{tikz,pgfplots}
\usepgfplotslibrary{fillbetween}
\usepgfplotslibrary{groupplots}
\usetikzlibrary{shapes.geometric}

\newcommand*{\inlineequation}[2][]{%
  \begingroup
    \refstepcounter{equation}%
    \ifx\\#1\\%
    \else
      \label{#1}%
    \fi
    \relpenalty=10000 %
    \binoppenalty=10000 %
    \ensuremath{%
      #2%
    }%
    ~\@eqnnum
  \endgroup
}
\makeatother
\usepackage{hyperref}

\usepackage{paralist}
\usepackage{xcolor}
\tikzset{%
	partial ellipse/.style args={#1:#2:#3}{%
		insert path={+ (#1:#3) arc (#1:#2:#3)}%
	}%
}%

\begin{document}
\selectlanguage{english}    

\title{Over-the-fiber Digital Predistortion\\ Using Reinforcement Learning}%



\author{
Jinxiang Song\textsuperscript{(1,*)}, Zonglong He\textsuperscript{(2,*)},
    Christian H\"{a}ger\textsuperscript{(1)}, Magnus Karlsson\textsuperscript{(2)},\\
    Alexandre Graell i Amat\textsuperscript{(1)},  Henk Wymeersch\textsuperscript{(1)}, and Jochen Schr\"{o}der\textsuperscript{(2)}
}
\maketitle                  


\setstretch{1.1}
\renewcommand\footnotemark{}
\renewcommand\footnoterule{}

\let\thefootnote\relax\footnotetext{978-1-6654-3868-1/21/\$31.00 \textcopyright 2021 IEEE}
\vspace{-2cm}
\begin{strip}
 \begin{author_descr}
 
   \textsuperscript{(1)}
   Dept.~of Electrical Engineering, Chalmers Univ.~of Technology, Gothenburg, Sweden\\
   \textsuperscript{(2)}
   Dept.~of Microtechnology and Nanoscience, Chalmers Univ.~of Technology, Gothenburg, Sweden\\
   \textsuperscript{(*)} Authors with equal contributions, \textcolor{blue}{\uline{jinxiang@chalmers.se}} \\ 
 \end{author_descr}
\end{strip}

\begin{strip}
  \begin{ecoc_abstract}
    We  demonstrate, for the first time, experimental over-the-fiber training of transmitter neural networks (NNs) using reinforcement learning. Optical back-to-back training of a novel NN-based digital predistorter outperforms \emph{arcsine}-based predistortion with up to 60\% bit-error-rate reduction. 
  \end{ecoc_abstract}
\end{strip}


\section{Introduction}
\begin{figure*}[t]
    \centering
     \vspace{-1cm}
    \includegraphics[width=0.85\textwidth]{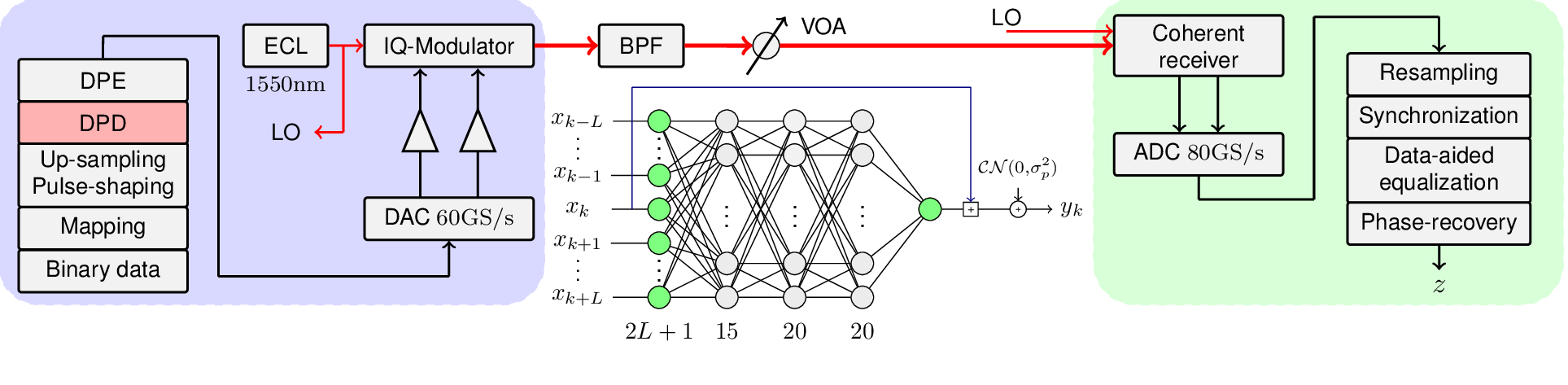}
    \vspace{-0.3cm}
    \caption{Experimental setup. Black lines: electrical path; Red lines: optical path; ECL: External cavity laser; BPF: bandpass filter; VOA: variable-optical attenuator. The inset figure shows the NN structure of the proposed DPD.}
    \label{fig:exp_setup}
    \vspace{0cm}
\end{figure*}

Modern high spectral efficiency  fiber-optic communication systems operate at high symbol rates with high-order constellation formats. However, cascaded linear and nonlinear  impairments induced by non-ideal hardware components including the digital-to-analog converter (DAC), power amplifier (PA), and in-phase-quadrature modulator (IQM), can severely degrade the system performance~\cite{khanna2015robust, berenguer2015nonlinear,
faig2019dimensions, elschner2018improving,yoffe2019low}. To improve the performance of optical communication systems, digital predistortion (DPD) is used to pre-compensate for the transmitter impairments.
The most commonly used DPD is based on linear digital pre-emphasis (DPE)~\cite{zhou2020impact, napoli2015novel, rafique2014digital} to compensate for the DAC frequency response  and is typically combined with  nonlinear \emph{arcsine}-based predistortion~\cite{curri2012optimization, tang2008coherent} to cancel the IQM nonlinearity. DPD based on Volterra series and
generalized
memory polynomials have also been widely studied for pre-compensation of radio-frequency amplifiers~\cite{eun1997new,kim2001digital,morgan2006generalized} and  optical transmitters~\cite{khanna2015robust, berenguer2015nonlinear,
faig2019dimensions, elschner2018improving,yoffe2019low}.

As an alternative to these model-based approaches, DPD based on neural-networks (NNs) has been studied in both wireless~\cite{ gotthans2014digital, benvenuto1993neural,tarver2019design,tarver2019neural}  and optical communications~\cite{Bajaj2020, paryanti2020direct, abu2019neural, paryanti2018recurrent}. Similar to its conventional counterparts, the NN-based DPD methods can be categorized into two groups, namely those using an indirect learning (ILA) and a direct learning architecture (DLA). 
It has been shown that, in general, DLA achieves better performance than ILA ~\cite{paryanti2018recurrent}. However, training an NN-based DPD applying DLA requires back-propagating the gradient over the nonlinear system, which is impossible in an experimental setting.
One way to circumvent this limitation is to develop a surrogate differentiable  model of the experimental channel and use this model for training~\cite{tarver2019neural,tarver2019design, Bajaj2020, paryanti2020direct,abu2019neural}. 
Similar approaches have been proposed in the context of end-to-end learning of transmitter--receiver pairs in both wireless~\cite{o2017introduction, dorner2017deep} and optical communications~\cite{karanov2018end,li2018achievable}. 
However, approaches using surrogate models require additional modeling efforts and may lead to performance degradation in cases where the model deviates significantly from the real system~\cite{dorner2017deep}.
A different approach to relieve the limitation of gradient back-propagation 
for transmitter learning is based on reinforcement learning (RL), complemented by supervised learning (SL) of the receiver~\cite{aoudia2019model}. 

In this paper, we demonstrate, for the first
time, training of transmitter NNs over an optical back-to-back channel using RL. Compared to ~\cite{aoudia2019model},
where transmitter optimization of signal constellations 
was demonstrated over memoryless 
channels, we focus on optimization of a NN-based DPD with memory input for the optical transmitter impaired with both dispersive and nonlinear effects. In contrast to other NN-based DPD employing DLA~\cite{Bajaj2020, paryanti2020direct, tarver2019neural,tarver2019design}, the proposed DPD is directly trained over the optical channel without any surrogate model. Experimental results show that the proposed DPD effectively mitigates nonlinear effects and achieves significantly better bit-error-rate (BER) performance than a baseline scheme with  DPE and \emph{arcsine}.

\section{The proposed DPD}
\noindent \emph{NN structure:}
The architecture of the proposed NN-based DPD is shown in the inset figure in Fig.~\ref{fig:exp_setup}. The DPD NN consists of an input layer, an output layer with linear activation, and $3$ fully-connected hidden layers with ReLU activation. The detailed number of neurons in each layer is also shown in Fig.~\ref{fig:exp_setup}.  Similar to ~\cite{mkadem2011physically, wu2020residual, gotthans2014digital}, in order to take the memory effect into account, the input layer is fed with $2L+1$ real-valued signals, where one signal corresponds to the current instantaneous input and the remaining $2L$ signals  to the inputs of the previous and future $L$ time steps, 
respectively. Moreover, a shortcut connection is employed to directly add the current input to the output, which has been shown to allow for better performance and quicker convergence rate~\cite{Bajaj2020,wu2020residual, tarver2019neural,tarver2019design}.

\noindent \emph{Data-aided digital signal processing (DSP) for training:}
To train the DPD, the oversampled signal received at the receiver needs to be sent back to the DPD. To do that, the ADC output is first resampled to have the same rate as the transmitted signal. Then, data-aided DSP, which is shown in Fig.~\ref{fig:exp_setup}, including time-synchronization, equalization, and phase-recovery is applied, after which the equalized oversampled signal $z$ is sent back to the DPD. In contrast to conventional DSP~\cite{mazur2019overhead} where a fractional-spaced equalizer is used, the data-aided DSP used here operates at the sample level, and is more complex to run in a real system. However, we note that such data-aided DSP is only needed for training the proposed DPD, and regular DSP is used after training.

\noindent \emph{DPD training:} Training of the proposed DPD is performed by modifying the approach in ~\cite{aoudia2019model}, with the goal to minimize the mean-square-error between the oversampled transmitted and received signals. Given the baseband pulse-shaped transmit signal $x_k$, the NN-based DPD, denoted by $f_{\theta}$, where $\theta$ is the set of trainable parameters,  takes a vector of $2L+1$ signals $\boldsymbol{x}_k=[x_{k-L},\ldots,x_k, \ldots x_{k+L}]$ as input and generates the pre-distorted signal to which a Gaussian perturbation is applied such that a random signal $y_k \sim \mathcal{CN}(f_{\theta}(\boldsymbol{x}_k), \sigma_p^2))$, where $\sigma_p^2$ is the variance of the Gaussian perturbation, is transmitted.
Then, the parameters of the NN-based DPD are updated according to 
\inlineequation[eq:gradientupdate]{\theta_{t+1} = \theta_t - \alpha \nabla_\theta \ell  (\theta_t) }, where $\nabla_\theta \ell (\theta) = \frac{1}{N} \sum_{k=1}^{N} |z_k - x_k|^2  \nabla_\theta  \log \pi_\theta ({y}_k | f_{\theta}(\boldsymbol{x}_k))$,  $\alpha$ is the learning rate, $N$ is the training batchsize and $\pi_\theta ({y}_k | f_{\theta}(\boldsymbol{x}_k)$ is the probability of transmitting $y_k$ given $f_{\theta}(\boldsymbol{x}_k)$. 
In our experiment, we set $\alpha=0.001$ and $N=2^{17}$.

\section{Experimental 
setup}
The experimental setup is shown in Fig.~\ref{fig:exp_setup}. Binary sequences are first mapped to constellation points and 2--times upsampled before pulse-shaped by a root-raised cosine filter with 10\% roll-off. The resulting signals are fed to the NN-based DPD, after which a DPE filter is applied to pre-compensate for the DAC frequency response. The purpose of using the DPE is to pre-compensate for most of the linear memory effects (e.g., the narrow DAC frequency response) so that the required DPD input length can be reduced~\textemdash the NN only needs to deal with the residual memory and nonlinear effects. The pre-compensated signals are then fed to a 60\,GS/s DAC (24\,GHz bandwidth) and amplified by an electrical PA (SHF-824, 35\,GHz) to drive the IQM. The optical carrier is generated by an external cavity laser (ECL) operating at 1550\,nm, and 50\% of the ECL output is sent to the coherent receiver for self-homodyne detection. Similar to ~\cite{Bajaj2020}, we consider a back-to-back setup, where the optical path includes a bandpass filter (BPF) and a variable-optical-attenuator (VOA). At the receiver, the signal is detected by a coherent receiver and sampled by a 80\,GS/s analog-to-digital converter (ADC). BER performance measurements use independent random data and pilot-based DSP~\cite{mazur2019overhead} to reconstruct the signal. 

\begin{figure*}[h]
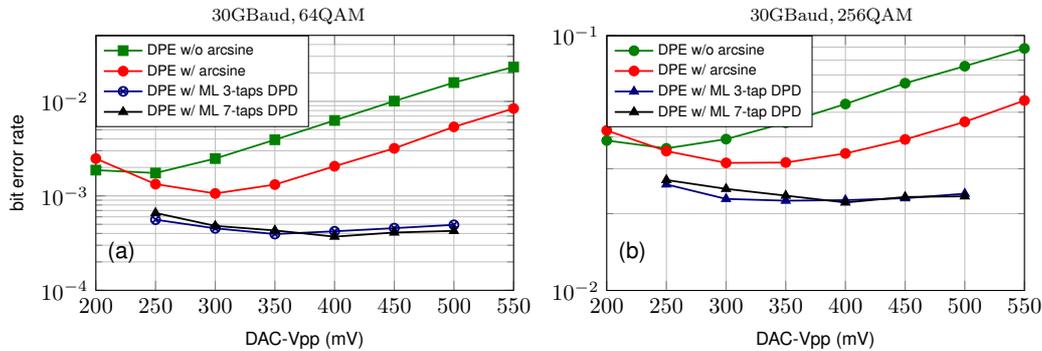

    \centering
    \vspace{-1.1cm}
    \includestandalone[width=0.86\textwidth]{figures/64qam_ber}
    \vspace{-1mm}
    \caption{BER performance versus DAC output voltage for (a): 64QAM and (b) 256QAM; The green and blue curve serve as two baseline schemes, where the green one applies linear DPE only while the blue one applies both DPE and \emph{arcsine} and clipping. For 256QAM, the DPD is trained with 64QAM signal, and directly applied to the 256QAM transmission.}
    \label{fig:ber_64}
\end{figure*}

\begin{figure}[t]
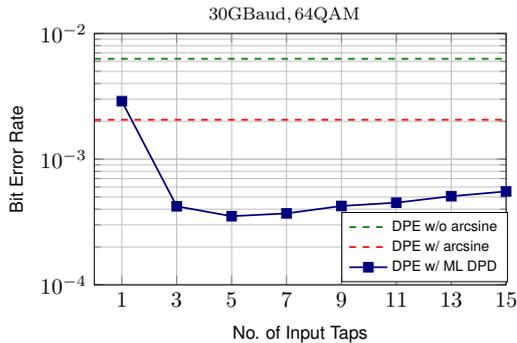

    \vspace{-2mm}
    \centering
    \includestandalone[width=0.9\columnwidth]{figures/ber_vs_mem}
     \vspace{-1mm}
    \caption{BER performance versus the DPD input length for 64QAM at DAC-Vpp 400\, mV.  
    Dashed curves: baselines with DPE (green) and DPE and \emph{arcsine} (red).}
    \label{fig:ber_vs_len_mem}
\end{figure}

We train the DPD by running the algorithm described in the previous section for 300 iterations. In each iteration, the DPD generates the pre-distorted signals and feeds them to the optical channel after the Gaussian perturbation is applied. At the receiver, data-aided DSP is applied to the received signals and the signal is sent back to the DPD and used for updating the DPD parameters according to~\eqref{eq:gradientupdate}. Finally, the DPD with updated parameters is used to generate the pre-distorted signals in the next training iteration.

\section{Results}
\label{label:results}
\noindent\emph{Baseline performance:} 
We start with measuring the performance of two commonly used baseline schemes. 
As a first baseline, a linear DPE scheme is applied to compensate for the DAC frequency response. An improved baseline combines DPE and \emph{arcsine} with clipping, where the \emph{arcsine} linearizes the modulator response while the optimized clipping reduces the peak-to-average power ratio. The achieved BER performance for 64QAM and 256QAM with 30\,GBaud transmission is shown in Fig.~\ref{fig:ber_64} (a) and (b), respectively.

\noindent\emph{BER performance on 64QAM:}
We now train and evaluate the performance of the proposed NN-based DPD in a 64QAM transmission.  We note that for each DAC output peak-to-peak voltage (DAC-Vpp), a separate DPD needs to be trained due to the fact that each DAC-Vpp corresponds to a different drive voltage swing, and therefore different output signal power as well as different transmitter nonlinear effects. Fig.~\ref{fig:ber_64} (a) visualizes the BER performance of the proposed DPD. For a range of DAC-Vpps, our approach, with input length set to $2L+1=3$ (blue) and $2L+1=7$ (black), achieves significant better BER performance than the baseline schemes. 

\noindent\emph{BER performance on 256QAM:}
We then apply the DPD trained for 64QAM to the 256QAM transmission without retraining. The achieved BER is shown in Fig.~\ref{fig:ber_64} (b). Interestingly, the DPD trained for 64QAM signals outperforms the two baselines for all considered DAC-Vpps, indicating that the proposed DPD has the flexibility to be trained for a single constellation format, but deployed for  different constellation formats.   

\noindent\emph{Impact of DPD input length:} 
We evaluate the impact of the DPD input length on the the BER performance. To have a fair comparison, all DPDs are trained for the same number of iterations (300 iterations) with the DAC-vpp set to 400\,mV. Fig.~\ref{fig:ber_vs_len_mem} shows the achieved BER performance as a function of the DPD input length. For a DPD input length $1$, it is shown that the proposed NN-based DPD achieves better performance than the baseline scheme applying DPE only, while performs slightly worse than the one using both DPE and \emph{arcsine} with optimized clipping. Such performance is expected, because the NN-based DPD should still be able to learn to compensate for the modulator response, similar to what the \emph{arcsine} does. As we increase the DPD input length, the BER performance of the proposed scheme improves. With the DPD input length set to $2L+1=5$, the optimal performance is achieved. However, when the input length is further increased, the BER increases. A possible explanation of this behavior is that, the NN-based DPD with large input length may require more training iterations, and training all NNs with the same number of iterations (i.g., 300) may lead to insufficient training for the large NNs.

\noindent\emph{DPD stability:} Finally, we evaluate the stability of the trained DPD. Our measurements show that the performance of the trained DPD remains stable over hours and can be repeated after several days without retraining. Such results seem to indicate that the proposed scheme requires rare retraining and may only need to be trained at link setup or transceiver calibration.

\section{Conclusions}
We demonstrated a novel over-the-fiber training of NN-based DPD utilizing reinforcement learning. Our experimental results show that the proposed DPD significantly outperforms the considered baseline schemes with up to 60\% BER reduction for a 30\,GBaud single channel transmission. 


\section{Acknowledgements}

This work was supported by the Knut and Alice Wallenberg Foundation, grant No.~2018.0090, and the Swedish Research Council under grant  No.~2018-0370.


\printbibliography

\end{document}

%% file: figures/64qam_ber.tex
\begin{tikzpicture}
\begin{semilogyaxis}[%
    scale only axis,
	xlabel= DAC-Vpp (mV) , ylabel= bit error rate,
	xmin=200, xmax=550,%
	ymin=1.e-4, ymax=5e-2,%
	grid=both,%
 	width=6.5cm, height=4.0cm,%
	legend cell align={left},%
	label style={font=\footnotesize},
	tick style={font=\footnotesize},
	legend style={at={(0,0.82)},anchor=west,nodes={scale=0.8, transform shape}, font=\footnotesize,  fill=white}%
]%

\addplot +[thick, color=green!50!black, solid, mark=square*, mark options={scale=1,solid }] coordinates { 
(150, 0.00404012  )
(200, 0.00187352) 
(250, 0.00174494 ) 
(300, 0.00248127)
(350, 0.00392603 )
(400,  0.0062853)
(450, 0.01007211)
(500,  0.01581394)
(550, 0.02311977  )
(600, 0.03239783)
};\addlegendentry{DPE w/o arcsine}

\addplot +[thick, color=red, solid, mark=*, mark options={scale=1,solid }] coordinates { 
(150, 0.0074783  )
(200, 0.00248118 ) 
(250, 0.00133183 ) 
(300, 0.00106011)
(350, 0.00131648 )
(400, 0.00205847)
(450, 0.00318565)
(500, 0.00537182)
(550, 0.00841187  )
(600, 0.01257405)
(650, 0.01878997)
};\addlegendentry{DPE w/ arcsine}

\addplot +[thick, color=blue!50!black, solid, mark=otimes, mark options={scale=1,solid }] coordinates { 

(250, 0.00055952 ) 
(300, 0.00045309)
(350, 0.00039428 )
(400, 0.00042159)
(450, 0.00045517)
(500, 0.00049334)
};\addlegendentry{DPE w/ ML 3-taps DPD}

\addplot +[thick, color=black, solid, mark=triangle*, mark options={scale=1,solid }] coordinates { 

(250, 0.00065907 ) 
(300, 0.00048109)
(350, 0.00043066 )
(400, 0.00037017)
(450, 0.00040928)
(500, 0.00042605)
};\addlegendentry{DPE w/ ML 7-taps DPD}

\end{semilogyaxis}
\node at (0.4, 0.6) {(a)};
\node at (3, 4.3) {\footnotesize{$30\mathrm{GBaud}, 64\mathrm{QAM}$}};
\end{tikzpicture}

\begin{tikzpicture}
\begin{semilogyaxis}[%
    scale only axis,
	xlabel= DAC-Vpp (mV) , 
	xmin=200, xmax=550,%
	ymin=1.0e-2, ymax=1e-1,%
	grid=both,%
 	width=6.5cm, height=4.0cm,%
	legend cell align={left},%
	label style={font=\footnotesize},
	tick style={font=\footnotesize},
	legend style={at={(0,0.82)},anchor=west,nodes={scale=0.8, transform shape}, font=\footnotesize},
]%

 \addplot +[thick, color=green!50!black, solid, mark=*, mark options={scale=1,solid }] coordinates { 
(150, 0.04941768 )
(200, 0.03872984 )
(250, 0.03607895)
(300, 0.03921842)
(350, 0.0454659)
(400, 0.05383373)
(450, 0.06496017)
(500, 0.07577483)
(550, 0.08894267)
(600, 0.10312671)

};\addlegendentry{DPE w/o arcsine}

\addplot +[thick, color=red, solid, mark=*, mark options={scale=1,solid }] coordinates { 
(150, 0.06036498 )
(200, 0.04241381 )
(250, 0.03520396)
(300, 0.03164037)
(350, 0.03174402)
(400, 0.03446127)
(450, 0.03911001)
(500, 0.0458301)
(550, 0.05558368)
(600, 0.06482546)
};\addlegendentry{DPE w/ arcsine}

\addplot +[thick, color=blue!50!black, solid, mark=triangle*, mark options={scale=1,solid }] coordinates { 
(250, 0.02606164)
(300, 0.02285445)
(350, 0.02249244)
(400, 0.02260129)
(450, 0.02299834)
(500, 0.02395576)
};\addlegendentry{DPE w/ ML 3-tap DPD}

\addplot +[thick, color=black, solid, mark=triangle*, mark options={scale=1,solid }] coordinates { 
(250, 0.02711181)
(300, 0.02502732)
(350, 0.02354546)
(400, 0.02213002)
(450, 0.02326093)
(500, 0.02339645)
};\addlegendentry{DPE w/ ML 7-tap DPD}

\end{semilogyaxis}

\node at (0.4, 0.6) {(b)};
\node at (3.5, 4.3) {\footnotesize{$30\mathrm{GBaud}, 256\mathrm{QAM}$}};
\end{tikzpicture}

%% file: figures/ber_vs_mem.tex
\begin{tikzpicture}
\begin{semilogyaxis}[%
    scale only axis,
	xlabel= No. of Input Taps  , ylabel= Bit Error Rate,
	xmin=0, xmax=15,%
	ymin=1.e-4, ymax=1e-2,%
	grid=both,%
 	width=6.5cm, height=4.0cm,%
	legend cell align={left},%
	label style={font=\footnotesize},
	tick style={font=\footnotesize},
	xtick={1,3,5,7,9,11,13,15},
	legend style={at={(0.6,0.15)},anchor=west,nodes={scale=0.8, transform shape}, font=\footnotesize}%
]%

\addplot+[thick, color=green!50!black, dashed, mark=none] coordinates { 
(0, 0.0062853)
(16, 0.0062853)
}; \addlegendentry{DPE w/o arcsine}

\addplot+[thick, color=red, dashed, mark=none] coordinates { 
(0, 0.00205847)
(16, 0.00205847)
}; \addlegendentry{DPE w/ arcsine}

\addplot +[thick, color=blue!50!black, solid, mark=square*, mark options={scale=1,solid }] coordinates { 
(1, 0.0028921033741858403  )
(3, 0.0004215925733225421) 
(5, 0.00035161820248969054 ) 
(7, 0.0003701656260839403)
(9, 0.0004238808918179366 )
(11,  0.00045052775850772727)
(13, 0.0005071335318148533)
(15,  0.0005534117624388176)

};\addlegendentry{DPE w/ ML DPD}

\end{semilogyaxis}
\node at (3, 4.3) {\footnotesize{$30\mathrm{GBaud}, 64\mathrm{QAM}$}};
\end{tikzpicture}